# Atomically-flat, chemically-stable, superconducting epitaxial thin film of iron-based superconductor, cobalt-doped BaFe$_2$As$_2$

Takayoshi Katase [a, *], Hidenori Hiramatsu [b], Hiroshi Yanagi [a], Toshio Kamiya [a, b], Masahiro Hirano [b, c], and Hideo Hosono [a, b, c] a: Materials and Structures Laboratory, Mailbox R3-1, Tokyo Institute of Technology, 4259 Nagatsuta-cho, Midori-ku, Yokohama 226-8503, Japan b: ERATO–SORST, Japan Science and Technology Agency, in Frontier Research Center, Tokyo Institute of Technology, S2-6F East, Mailbox S2-13, 4259 Nagatsuta-cho, Midori-ku, Yokohama 226-8503, Japan c: Frontier Research Center, S2-6F East, Mailbox S2-13, Tokyo Institute of Technology, 4259 Nagatsuta-cho, Midori-ku, Yokohama 226-8503, Japan

**Abstract**

Epitaxial growth of Fe-based superconductors such as Co-doped $SrFe_2As_2$ ($SrFe_2As_2$:Co) was reported recently, but has still insufficient properties for device application because they have rough surfaces and are decomposed by reactions with water vapor in an ambient atmosphere. This letter reports that epitaxial films of Co-doped $BaFe_2As_2$ grown at 700 °C show the onset superconducting transition tempearture of ~20 K. The transition is sharper than those observed on the $SrFe_2As_2$:Co films, which would originate from their improved crystallinity. These films also have atomically-flat surfaces with steps-and-terraces structures and exhibit chemical stability against exposure to water vapor.

----------------------------------------

Footnotes:

[*] Electronic mail: katase@lucid.msl.titech.ac.jp

Extensive researches on the superconductivity in Fe-based layered compounds have been continuing since the discovery of superconductivity at 26 K in F-doped LaFeAsO,[1] and the highest superconducting transition temperature ($T_c$) over 50 K in $RE$FeAsO ($RE$ = rare earths)[2,3] has been reported. New Fe-based high-$T_c$ superconducting materials have been found also in other systems containing a square lattice of Fe atoms, which include $AE$Fe$_2$As$_2$ ($AE$ = alkaline earths),[4] $A$FeAs ($A$ = alkalis),[5] $AE$FeAsF,[6] FeSe,[7] and perovskite oxide layers-containing iron-pnictides e.g. Sr$_4$Sc$_2$O$_6$Fe$_2$P$_2$.[8] A prominent feature of the Fe-based superconductors is their high critical magnetic fields,[9,10] which are important for practical applications.

To measure intrinsic properties of the Fe-based superconductors and apply them to superconductor devices, it is important to grow high-quality epitaxial films. We reported that pulsed laser deposition (PLD) excited by a Nd:YAG second harmonic laser using a high-purity target produced epitaxial films of LaFeAsO,[11] and the modified PLD method is also applicable to Co-doped SrFe$_2$As$_2$ (SrFe$_2$As$_2$:Co).[12] The SrFe$_2$As$_2$:Co epitaxial films showed a superconducting transition at 20 K with a small magnetic field anisotropy.[13] After that, thin film growth of LaFeAsO[14] and Fe(Se, Te)[15-19] have been reported by other groups.

A drawback of the resulting SrFe$_2$As$_2$:Co thin films is insufficient chemical stability in an ambient atmosphere; i.e., they have somewhat high reactivity with water. We recently found that this instability to water vapor led to unexpected emergence of superconductivity at 25 K in undoped SrFe$_2$As$_2$.[20] Although the $T_c$ of the water-exposed SrFe$_2$As$_2$ epitaxial films is higher than that of the SrFe$_2$As$_2$:Co epitaxial films, the water-exposed films produced a large amount of impurities such as Fe$_2$As. Therefore, the high sensitivity to water vapor of undoped and Co-doped SrFe$_2$As$_2$ [SrFe$_2$As$_2$(:Co)] films is an obstacle for practical use and application to superconductor devices such as Josephson junctions. In addition, the epitaxial films of LaFeAsO and $SrFe_2As_2$(:Co) have rough surfaces with granular grains, which is thought to be an origin of the low critical current density (~20 kA/cm$^2$) of $SrFe_2As_2$:Co superconducting epitaxial films[21] and is another obstacle for fabricating superconductor devices.

In this letter, we report growth of superconducting epitaxial film of Co-doped $BaFe_2As_2$ ($BaFe_2As_2$:Co) with a $T_c$ similar to that of bulk samples.[22] Contrary to $SrFe_2As_2$(:Co), the $BaFe_2As_2$:Co films did not exhibit detectable deterioration in film / crystal structures, impurity phases, and superconducting properties. The $BaFe_2As_2$:Co films had better crystallinity than $SrFe_2As_2$(:Co), and their surfaces were atomically-flat although droplets and pits were found in large-area observations.

$AE$Fe$_2$As$_2$:Co ($AE$ = Sr, Ba) thin films (thickness: 500 nm) were fabricated on mixed perovskite (La,Sr)(Al,Ta)O$_3$ (LSAT) (001) single-crystal substrates by ablating polycrystalline bulk targets of $AE$Fe$_{1.8}$Co$_{0.2}$As$_2$ using a second harmonic (wavelength: 532 nm) of a pulsed Nd:YAG laser in a vacuum at ~10$^{-5}$ Pa. The growth temperature ($T_g$) was fixed at 700 $^o$C. The targets were synthesized by a solid state reaction of stoichiometric mixtures of $AE$As, Fe$_2$As, and Co$_2$As via a reaction $AE$As + 0.9 Fe$_2$As + 0.1 Co$_2$As → $AE$Fe$_{1.8}$Co$_{0.2}$As$_2$.[23] The mixtures were pressed into disks and then heated in evacuated silica tubes at 900 $^o$C for 16 hours.

Film structures including crystalline quality and crystallites orientation were examined by high-resolution x-ray diffraction (HR-XRD, anode radiation: CuK$\alpha_1$, ATX-G, Rigaku) at room temperature (RT). The film surface morphology was observed in an ambient air at RT with an atomic force microscope (AFM, SPI-3800N, S.I.I.). The temperature ($T$) dependence of electrical resistivity ($\rho$) was measured by the four-probe method using Au electrodes in the temperature range of 2 – 305 K with a physical property measurement system (PPMS, Quantum Design). External magnetic fields ($H$) of 0 – 9 T were applied during the $\rho - T$ measurements. Chemical stability against water vapor was examined by comparing the XRD and electrical measurement results between samples before and after exposure to water vapor. The exposure tests were carried out at RT in a wet $N_2$ flow with the dew point of 20 – 25 ºC generated from distilled water. The exposure time was varied up to 48 hours.

Figure 1(a) shows an out-of-plane HR-XRD pattern of the $BaFe_2As_2$:Co film. It shows intense peaks assigned to $BaFe_2As_2$:Co 00$l$ and LSAT 00$l$ diffractions. An extra peak was also observed at $2\theta = $ ~65 degrees (indicated by the vertical arrow), which can be assigned to the 200 diffraction of Fe metal. This obsrvation indicates that the $BaFe_2As_2$:Co film was $c$-axis oriented strongly, and did not contain the FeAs impurity phase that was detected in the $SrFe_2As_2$:Co epitaxial films at $2\theta = $ ~35 degrees.[12] In addition, the FWHM ($\Delta\omega$) of the $BaFe_2As_2$:Co 002 rocking curve ($2\theta$ fixed $\omega$ scan, inset of Fig. 1(a)) was 0.6 degrees, which was approximately a half of that of the $SrFe_2As_2$:Co films (1.0 degrees). These results indicate that the $BaFe_2As_2$:Co films had higher quality than the $SrFe_2As_2$:Co films. Figure 1(b) shows the in-plane HR-XRD pattern and the in-plane $BaFe_2As_2$:Co 200 rocking curve ($2\theta_\chi$ fixed $\phi$ scan, inset). The in-plane HR-XRD pattern shows $BaFe_2As_2$:Co 200 and LSAT 400 diffractions along with a diffraction from the impurity Fe phase (Fe 110), which is indicated by the vertical arrow at $2\theta_\chi = $ ~45 degrees. Note that we observed a similar in-plane diffraction peak also in the $SrFe_2As_2$:Co epitaxial films, but we assigned it to FeAs in ref. 12. Taking the above results for the $BaFe_2As_2$:Co films into consideration, we think that the impurity phase in the $SrFe_2As_2$:Co films would also be Fe metal, but it is difficult to conclude definitively because no other diffraction peak assignable to the impurity was detected. The $\phi$ scan of the BaFe$_2$As$_2$:Co 200 diffraction in inset shows a four-fold rotational symmetry originating from the tetragonal symmetry of the BaFe$_2$As$_2$:Co lattice.[24] These results substantiate that the BaFe$_2$As$_2$:Co film was grown heteroepitaxially on the LSAT (001) substrate with the epitaxial relationship of [001] BaFe$_2$As$_2$:Co // [001] LSAT for out-of-plane and [100] BaFe$_2$As$_2$:Co // [100] LSAT for in-plane. The lattice parameters of the obtained BaFe$_2$As$_2$:Co epitaxial film were $a$ = 0.3962 nm and $c$ = 1.296 nm, which are slightly smaller (–0.06% for the $a$-axis and –0.19% for the $c$-axis) than those of a bulk single crystal of BaFe$_{1.8}$Co$_{0.2}$As$_2$ ($a$ = 0.39639 nm and $c$ = 1.2980 nm[22]).

Figures 1(c) – (e) show the AFM images of a SrFe$_2$As$_2$:Co (c) and a BaFe$_2$As$_2$:Co epitaxial film (d, e), respectively. The SrFe$_2$As$_2$:Co epitaxial film had the granular structure with the average grain size of ~300 nm, and its surface was rather rough with the root mean-square roughness ($R_{rms}$) of 17.6 nm. On the other hand, step-flow growth was observed for the BaFe$_2$As$_2$:Co epitaxial film (e) as evidenced by the steps-and-terraces structure, although droplets and pits were observed in a larger area observeation (d). The cross-sectional height profile intersecting the line A–B in (e) reveals that the BaFe$_2$As$_2$:Co film had an atomically-flat terraces-and-steps surface with steps ~50 nm in width and 1.3 nm in height; this step height agrees with the $c$-axis length of the unit cell.

Figure 2 shows the $\rho - T$ curves and the effects of applied magnetic field ($H$) on the superconducting transition of the BaFe$_2$As$_2$:Co epitaxial film. $H$ was applied parallel to the $c$-axis in (a) ($H_{//c}$) and the $a$-axis in (b) ($H_{//a}$) of the BaFe$_2$As$_2$:Co lattice in the film. The RT resistivity of the BaFe$_2$As$_2$:Co film was $\rho_{300K}$ = 1.3 × 10$^{-3}$ $\Omega$cm, which is much larger than that of a single crystal (~$3.2 \times 10^{-4}$ Ωcm). The onset $T_c$ ($T_c^{onset}$) and the offset $T_c$ ($T_c^{offset}$) were 20 K and 17 K, respectively. It gives the transition width at $H = 0$ T ($\Delta T_c = T_c^{onset} - T_c^{offset}$) of ~3 K, which is a half of those of the SrFe$_2$As$_2$:Co epitaxial films ($\Delta T_c$ = 6 K)[12]. The $T_c^{onset}$ of the BaFe$_2$As$_2$:Co film is slightly lower by 2 K than that of BaFe$_2$As$_2$:Co single-crystal ($T_c^{onset}$ of 22 K and $\Delta T_c$ of 0.6 K are reported[22]).

Under magnetic fields, $T_c^{onset}$ and $T_c^{offset}$ decreased and $\Delta T_c$ increased with the increasing $H$. The $T_c^{onset}$, $T_c^{offset}$, and $\Delta T_c$ at 9 T were 17.0 K, 11.4 K, 5.6 K for $H_{//c}$ and 19.0 K, 13.5 K, 5.5 K for $H_{//a}$, respectively. At temperatures ≤ 11.4 K for $H_{//c}$ and ≤ 13.5 K for $H_{//a}$, the superconducting state survived even when $H$ was raised up to 9 T, indicating that the upper critical magnetic fields ($H_{c2}$) are much higher than 9 T. The superconducting state was more sensitive to $H_{//c}$ than to $H_{//a}$, but the magnetic anisotropy was as small as that of the SrFe$_2$As$_2$:Co epitaxial films.[12]

Figure 3 summarizes the chemical stability against water vapor in comparison with the SrFe$_2$As$_2$:Co films. Figures 3(a,b) show the out-of-plane HR-XRD patterns around the 002 diffraction of the SrFe$_2$As$_2$:Co (a) and the BaFe$_2$As$_2$:Co (b) epitaxial films exposed to water vapor for 0 – 24 hours. The virgin films were measured immediately after the samples were taken out from the deposition chamber. In the case of the SrFe$_2$As$_2$:Co epitaxial films (Fig. 3(a)), when exposed for 2 h, the 002 diffraction peak broadened and an impurity phase assigned to Fe$_2$As appeared at $2\theta$ ~14.9 degrees (the vertical arrow). With increasing the exposure time, the SrFe$_2$As$_2$:Co 002 diffraction peak weakened, broadened, and up-shifted, which was accompanied by the increase in the Fe$_2$As impurity phase. These observations are very similar to those of the undoped SrFe$_2$As$_2$ epitaxial films.[20] Figures 3(c,d) show the exposure time dependences of the $c$-axis lattice parameters (c) and the FWHM of the 002 diffractions (d). These show that the $c$-axis length of the SrFe$_2$As$_2$:Co film was decreased from 1.233 nm for the virgin film to 1.225 nm, and the FWHM of the 002 diffraction increased from 0.17 to 0.57 degrees after the 24 hours of exposure. On the other hand, for the BaFe$_2$As$_2$:Co epitaxial film, very small changes were observed only when the exposure time was < 2 hours, and no observable change was detected even at prolonged exposure time. Also in the XRD pattern (Fig. 3(b)), no detectable change, including a symptom of an impurity phase, was observed.

The chemical stability of the BaFe$_2$As$_2$:Co films is confirmed also by the superconducting properties. Figures 3(e,f) show the $\rho - T$ curves of the SrFe$_2$As$_2$:Co (e) and the BaFe$_2$As$_2$:Co (f) epitaxial films as a function of exposure time to water vapor. The $T_c^{onset}$ of the SrFe$_2$As$_2$:Co films remained unchanged with increasing the exposure time, but the $T_c^{offset}$ obviously decreased with increasing the exposure time, which would result from the decrease in the volume fraction of the SrFe$_2$As$_2$:Co phase as observed in Fig. 3(a). In addition, the exposure to water vapor raised the RT resistivity from $3.8 \times 10^{-4}$ $\Omega$cm for the virgin film to $7.5 \times 10^{-4}$ $\Omega$cm after 48 hours of exposure. On the contrary, the BaFe$_2$As$_2$:Co epitaxial films were very stable against the water vapor exposure and did not exhibit any change in $T_c^{onset}$ and $T_c^{offset}$ although the slight increase in the RT resistivity was observed. These observations substantiate that the BaFe$_2$As$_2$:Co epitaxial films were much stable to water vapor than the SrFe$_2$As$_2$:Co epitaxial films. For bulk polycrystal samples, better stability of BaFe$_2$As$_2$:Co than SrFe$_2$As$_2$:Co is referred to in refs. 4, 24, 25. The preset results provide solid evidences for the chemical stability of BaFe$_2$As$_2$:Co.

In summary, we fabricated superconducting epitaxial films of BaFe$_2$As$_2$:Co, which exhibited the onset superconducting transition temperature of 20 K. The $BaFe_2As_2$:Co epitaxial films had higher crystallinity than those of the $SrFe_2As_2$:Co films. Their surfaces were basically composed of atomically-flat steps-and-terraces structures. What is more important is that the $BaFe_2As_2$:Co epitaxial films are very stable against water vapor at least for 48 hours of exposure at RT. The growth of the high crystallinity, atomically-flat epitaxial films and the high chemical stability of the $BaFe_2As_2$:Co epitaxial films will provide a large contribution to developing superconductor devices based on the Fe-based high-$T_c$ superconductors.

Phys.: Condens. Matter **20**, 452201 (2008).

**Figures**

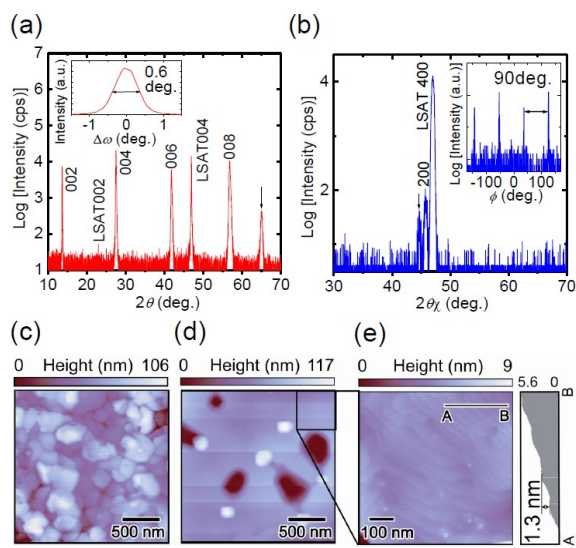

Fig. 1. (Color online) (a,b) Out-of-plane (a) and in-plane (b) HR-XRD patterns of the BaFe$_2$As$_2$:Co film grown at 700 °C. Insets show the rocking curves of the 002 diffraction (a) and the 200 diffraction (b), respectively. (c,d) Wide scan AFM images of SrFe$_2$As$_2$:Co (c) and BaFe$_2$As$_2$:Co (d) epitaxial films. The scan area is 2.0×2.0 μm$^2$. (e) Magnified image (0.7×0.7 μm$^2$) of the flat region in (d) (left) and cross-sectional height profie intersecting the line A–B (right).

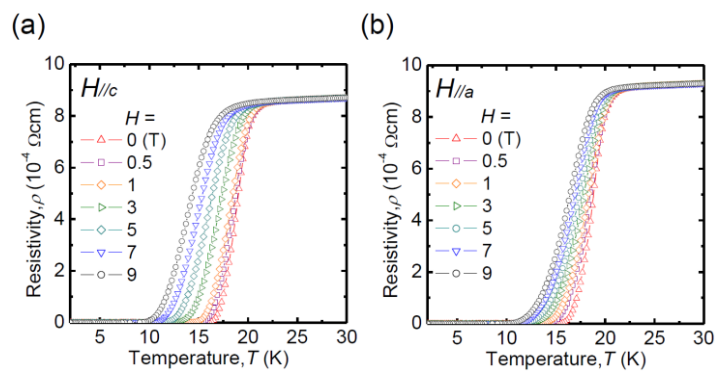

Fig. 2. (Color online) Temperature dependences of electrical resistivity ($\rho - T$) under magnetic fields ($H$) varying from 0 to 9 T applied parallel to the $c$-axis (a) and the $a$-axis (b) of the BaFe$_2$As$_2$:Co epitaxial film.

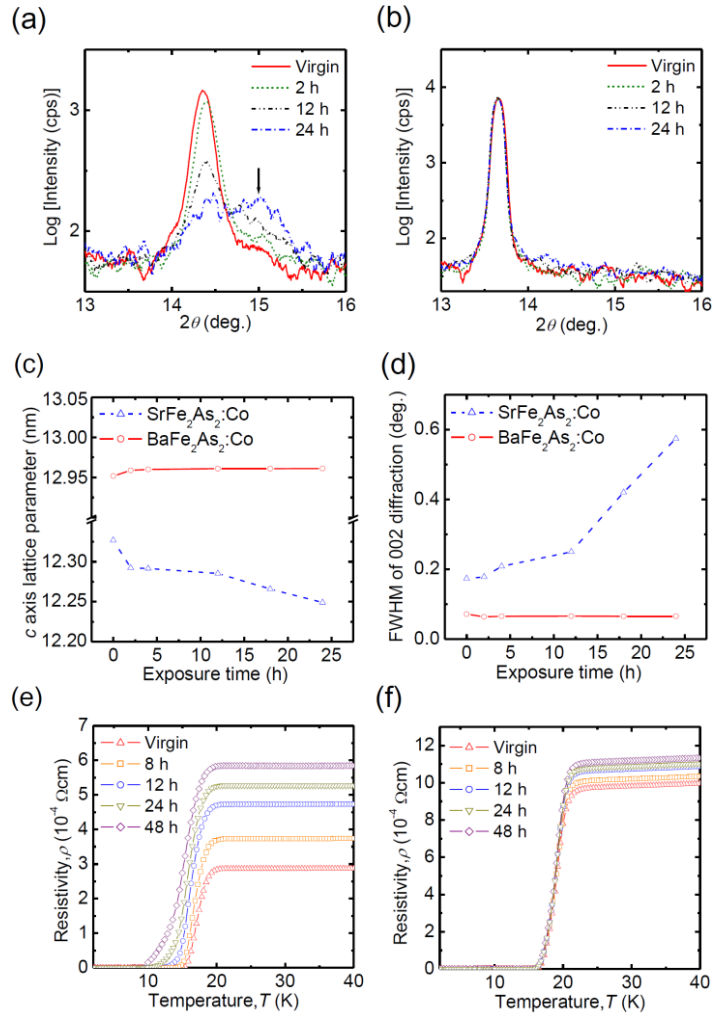

Fig. 3. (Color online) Effects of water vapor exposure at RT with dew point of 20–25 °C for varied exposure time up to 48 hours. (a,b) Changes in the out-of-plane HR-XRD patterns around the 002 diffractions of SrFe$_2$As$_2$:Co (a) and BaFe$_2$As$_2$:Co (b) epitaxial films. (c,d) Exposure time dependences of the *c*-axis lattice parameters (c) and the FWHMs of the 002 diffractions (d) of *AE*Fe$_2$As$_2$:Co (*AE* = Sr, Ba) epitaxial films. (e,f) Changes in the temperature dependences of electrical resistivity of SrFe$_2$As$_2$:Co (e) and BaFe$_2$As$_2$:Co (f) epitaxial films as a function of exposure time.